\documentclass[conference]{IEEEtran}
\IEEEoverridecommandlockouts
\usepackage{amsmath}
\usepackage{amssymb}
\usepackage{graphicx}
\usepackage{booktabs}
\usepackage{array}
\usepackage{tabularx}
\usepackage[table]{xcolor}
\usepackage{tikz}
\usetikzlibrary{arrows.meta,positioning,calc,fit}
\usepackage{cite}
\usepackage{balance}
\usepackage{url}

\definecolor{TitleBlue}{HTML}{173B63}
\definecolor{AccentBlue}{HTML}{2E5EAA}
\definecolor{AccentTeal}{HTML}{0F766E}
\definecolor{HeaderFill}{HTML}{DCEAF7}
\definecolor{NodeAMF}{HTML}{EAF2FC}
\definecolor{NodeBCEE}{HTML}{FFF4E7}
\definecolor{NodeNRF}{HTML}{EDF8EE}
\definecolor{NodeHybrid}{HTML}{E8F7F5}
\definecolor{BorderBlue}{HTML}{6D8FB5}
\definecolor{SoftGray}{HTML}{5B6573}

\newif\ifblindreview
\blindreviewfalse

\newcommand{\HybridGroup}{\texttt{X25519MLKEM768}}
\newcommand{\BCEEA}{\ensuremath{\mathrm{BCEE}_{\mathrm{A}}}}
\newcommand{\BCEEB}{\ensuremath{\mathrm{BCEE}_{\mathrm{B}}}}
\newcommand{\AMFTML}{\texttt{amf-tml}}
\newcommand{\NRFTML}{\texttt{nrf-tml}}
\newcommand{\SingleRunID}{\texttt{18146786-\allowbreak9fdf-\allowbreak41f1-\allowbreak9a89-\allowbreak3faae49c6d25}}

\newcommand{\COneMean}{26.502}
\newcommand{\COneSD}{10.028}
\newcommand{\COneMedian}{22.968}
\newcommand{\COneQOne}{20.174}
\newcommand{\COneQThree}{30.039}
\newcommand{\COnePNinetyFive}{46.993}
\newcommand{\COneMin}{9.395}
\newcommand{\COneMax}{58.503}
\newcommand{\CTwoMean}{30.481}
\newcommand{\CTwoSD}{10.286}
\newcommand{\CTwoMedian}{26.351}
\newcommand{\CTwoQOne}{23.465}
\newcommand{\CTwoQThree}{35.331}
\newcommand{\CTwoPNinetyFive}{48.565}
\newcommand{\CTwoMin}{13.586}
\newcommand{\CTwoMax}{61.221}
\newcommand{\MedianDelta}{3.383}
\newcommand{\MedianRelativeDelta}{14.73\%}
\newcommand{\MedianBlockCI}{[1.193, 7.872]}
\newcommand{\MedianBlockCINinetyNine}{[0.606, 9.088]}

\newcommand{\CliffsDelta}{0.292}
\newcommand{\BlockMWU}{31}
\newcommand{\BlockMWP}{0.0411}
\newcommand{\STwoCOneMedian}{10.652}
\newcommand{\STwoCTwoMedian}{15.251}
\newcommand{\STwoDelta}{4.599}
\newcommand{\STwoRelativeDelta}{43.2\%}
\newcommand{\STwoCI}{[3.294,~6.109]}
\newcommand{\STwoCINinetyNine}{[2.959,~6.278]}
\newcommand{\STwoBlockMWU}{36}
\newcommand{\STwoBlockMWP}{0.0022}
\newcommand{\STwoCOneBlockRange}{10.12--11.74}
\newcommand{\STwoCTwoBlockRange}{13.61--16.87}

\begin{document}
\setlength{\parskip}{0pt}

\title{Post-Quantum Migration in Brownfield 5G Core Networks: Verified Hybrid Key Establishment, Performance, and Rollback%
\thanks{\copyright{} 2026 IEEE. Personal use of this material is permitted.
Permission from IEEE must be obtained for all other uses, in any current
or future media, including reprinting/republishing this material for
advertising or promotional purposes, creating new collective works,
for resale or redistribution to servers or lists, or reuse of any
copyrighted component of this work in other works.}}

\ifblindreview
\author{\IEEEauthorblockN{Anonymous Author(s)}}
\else
\author{
\IEEEauthorblockN{Shubham Verma}
\IEEEauthorblockA{
Centre for Development of Telematics (C-DOT)\\
New Delhi, India
}
}
\fi

\maketitle

\begin{abstract}
Migrating brownfield 5G Core networks to post-quantum cryptography
(PQC) is difficult because deployed network functions, base images,
and TLS libraries evolve under different upgrade constraints. This paper
evaluates a deployment-controlled migration strategy that enables hybrid
post-quantum/traditional key establishment through externally deployed
Boundary Cryptographic Execution Entities (BCEEs), without modifying
participating network function source code. The approach is evaluated
using the Open5GS AMF--NRF Service-Based Interface as a representative
5G Core scenario. Classical and hybrid configurations are compared under
an unchanged mediation topology. Packet-level analysis confirms successful
negotiation of X25519MLKEM768, and functional validation demonstrates
that representative SBI procedures remain operational through the
mediated path. In 120 NF-registration transactions, the median AMF-facing
registration latency increased from 22.968~ms under the classical state
to 26.351~ms under the hybrid state, reflecting a 3.383~ms
(14.73 percent) increase, with a whole-block 95 percent confidence
interval of [1.193, 7.872]~ms. At the mediated TLS boundary, the hybrid configuration shows a
4.599~ms higher median segment-local TLS-establishment time than
the classical configuration. These two reported differences reflect
system-level runtime effects and do not isolate key-encapsulation cost. Packet
evidence verifies rollback from hybrid-mediated operation to classical
mediated operation and, separately, restoration of the native direct path.
NF-facing compatibility legs remained plaintext and outside the hybrid
post-quantum scope. The results demonstrate a packet-verifiable and
manually reversible brownfield mechanism for introducing hybrid
post-quantum key establishment on a controlled 5G SBI segment.
\end{abstract}\vspace{4pt}

\begin{table*}[!t]
\centering
\caption{Comparison of Closely Related 5GC PQC Approaches}
\label{tab:related_comparison}

\footnotesize
\renewcommand{\arraystretch}{1.00}
\setlength{\tabcolsep}{2.2pt}

\begin{tabularx}{\textwidth}{
    @{}
    >{\raggedright\arraybackslash}p{0.17\textwidth}
    >{\centering\arraybackslash}p{0.12\textwidth}
    >{\centering\arraybackslash}p{0.12\textwidth}
    >{\centering\arraybackslash}p{0.14\textwidth}
    >{\centering\arraybackslash}p{0.13\textwidth}
    >{\centering\arraybackslash}p{0.11\textwidth}
    >{\centering\arraybackslash}X
    @{}
}
\toprule

\textbf{Work} &
\shortstack{\textbf{External}\\\textbf{crypto}\\\textbf{boundary}} &
\shortstack{\textbf{NF code}\\\textbf{unchanged}} &
\shortstack{\textbf{RFC~10024}\\\textbf{TLS}\\\textbf{hybrid}} &
\shortstack{\textbf{Packet TLS}\\\textbf{evidence}} &
\shortstack{\textbf{Rollback}\\\textbf{shown}} &
\shortstack{\textbf{Functional}\\\textbf{5GC/SBI}} \\

\midrule

Attema et al.
& No & No & No & No & No & Yes \\

Scalise et al.
& No & No & No & Yes & No & Yes \\

Faval et al.
& Yes & Yes & No & No & No & Yes \\

Koranga
& No & No & No & No & No & Yes \\

QORE
& No & No & No & Yes & No & Yes \\

\textbf{This work}
& \textbf{Yes}
& \textbf{Yes}
& \textbf{Yes}
& \textbf{Yes}
& \textbf{Yes}
& \textbf{Yes} \\

\bottomrule
\end{tabularx}

\vspace{2pt}
\begin{minipage}{0.96\textwidth}
\footnotesize
\raggedright
``No'' indicates that the corresponding property is not established
in the cited work and does not necessarily mean that its implementation
lacks that capability. ``RFC~10024 TLS hybrid'' denotes explicit use of a
PQ/T hybrid group standardized in the published RFC. ``Packet TLS evidence''
denotes packet-capture-derived evidence of TLS handshake fields, including
the negotiated group. QORE evaluates X25519MLKEM768 using the precursor
IETF draft rather than the published RFC~10024.
\end{minipage}

\end{table*}


\section{Introduction}
The 5G Core (5GC) Service-Based Architecture (SBA)
depends on HTTP/2-based Service-Based Interfaces (SBIs)
between Network Functions (NFs) [1]--[3]. Moving these
interfaces toward post-quantum cryptography (PQC) is difficult
in brownfield networks because cryptographic libraries, vendor
releases, certification baselines, interoperability, and maintenance
windows rarely advance at the same pace. ML-KEM is standardized
in FIPS~203 [4], and RFC~10024 defines TLS~1.3
post-quantum/traditional (PQ/T) hybrid groups including
X25519MLKEM768, which combines X25519 with ML-KEM-768
[5]. This study selects X25519MLKEM768 (NamedGroup~4588,
0x11EC) rather than the two NIST-curve alternatives in RFC~10024.
Among these three hybrid groups, only X25519MLKEM768 is marked
\texttt{Recommended=Y} in the IANA TLS Supported Groups registry;
the two SecP-curve groups are marked \texttt{Recommended=N} and
address use cases requiring FIPS-approved mechanisms~[5].
RFC~10024 describes X25519MLKEM768 as often the most practical
choice for a single PQ/T hybrid combiner~[5], motivating its
selection for this brownfield study. The practical challenge is finding a
way to introduce such a cryptographic profile while some network
functions are still unable to support it natively. This is consistent
with broader crypto-agility guidance emphasizing cryptographic
inventory and controlled migration [6], [7], as well as international
and national quantum-safe migration initiatives [8], [9].
Table~\ref{tab:related_comparison} positions this approach against closely
related 5GC PQC work.

We evaluate a deployment-controlled boundary on one Open5GS
AMF--NRF path. Two Boundary Cryptographic Execution Entities
(BCEEs) preserve the observed NF-facing SBI behavior in this
deployment while placing the evaluated TLS key-establishment
profile under deployment control on the inter-BCEE segment $S_2$,
without modifying the participating NF cryptographic stacks. The
evaluation considers three states. C0 represents the native
AMF--NRF communication path without BCEE mediation. C1
represents the BCEE-mediated topology in which $S_2$ uses classical
TLS~1.3 with X25519 key establishment. C2 retains the same
BCEE-mediated topology as C1 but uses X25519MLKEM768
for hybrid PQ/T key establishment on $S_2$. C1 and C2 therefore share the same mediated topology and
OQS-capable software/provider stack; their state-specific
$S_2$ TLS configurations select X25519 and
X25519MLKEM768, respectively. 

The study asks four questions: RQ1, whether real SBI procedures
remain functional without participating NF source-code modification;
RQ2, whether packet evidence verifies the negotiated hybrid state
for a real SBI transaction and the expected negotiated group across
the performance captures; RQ3, whether C1/C2 timing separation
appears at the complete NF-registration transaction boundary and
at the controlled $S_2$ TLS-establishment boundary when both states
use the same BCEE topology; and RQ4, whether the deployment can
be manually reversed from C2 to C1 and, in a separate transition,
from C1 to native C0. C0 remains an operational path state used for reversal validation
and is not part of the final C1/C2 performance inference. Hybrid PQ/T key establishment is claimed only for $S_2$,
not for the complete AMF--NRF path.

The experimental contribution is fourfold. First, one synchronized
C2 capture correlates a real AMF NF-registration transaction with
a corresponding inter-BCEE TLS session and directly verifies
NamedGroup~4588 with 1216/1120-byte client/server key shares.
Second, PUT, POST, GET, and PATCH NF-management procedures
complete through the hybrid-mediated path. Third, a pre-specified balanced 12-block
sequence uses six separately initialized C1 blocks and six
separately initialized C2 blocks, with ten fresh registrations per
block (120 protocol-valid observations, all retained). At the
complete registration boundary, the C2--C1 median difference is
+3.383~ms (14.73\%) with a whole-block 95\% CI [1.193, 7.872]~ms.
At the controlled $S_2$ TLS-establishment boundary, the C2--C1
median difference is +4.599~ms. Fourth, transition captures verify
two manual reversals, C2$\rightarrow$C1 and separately
C1$\rightarrow$C0, with successful SBI procedures observed after
each reversal.

\section{Background and Related Work}

NRF NF-management procedures are specified in 3GPP
TS~29.510~[10], TLS~1.3 in RFC~9846~[11], and RFC~10024
assigns NamedGroup~4588 (0x11EC) to X25519MLKEM768~[5].
3GPP TR~33.938 provides a cryptographic inventory across
the 5G System, while TR~33.703 studies transition to PQC~[12],
[13]. GSMA PQ.03 further describes migration considerations
for legacy systems, including reliance on upgradeable components
or tunneling through a quantum-safe channel when some components
cannot be upgraded~[14]. These considerations motivate a
deployment-controlled cryptographic boundary whose lifecycle can
be decoupled from the participating NF application.

Prior 5GC PQC research includes NF-integrated approaches~[15],
[16] and externally mediated approaches such as Faval et al.~[17].
Faval et al. use external sidecar proxies in free5GC, preserving
participating NF application code while introducing ML-KEM-768 and
ML-DSA-based protection through the proxy layer.

The BCEE role is distinct from the 3GPP Service Communication
Proxy (SCP), which supports service communication and routing in
SBA deployments~[2]. In this work, the BCEEs act as TLS termination
boundaries that introduce and expose the negotiated cryptographic
state on $S_2$. A production implementation could place this capability
near an SCP or integrate it into the SCP data plane, but such integration
is outside the scope of the present prototype.

The present work addresses a different migration objective. It studies
deployment-controlled transition between classical and standardized
TLS~1.3 PQ/T hybrid cryptographic states on a fixed mediated topology.
It experimentally verifies X25519MLKEM768 as RFC~10024
NamedGroup~4588 on the wire, associates the negotiated TLS state
with a real SBI transaction, evaluates C1/C2 performance under the
same BCEE topology, and demonstrates packet-observed C2$\rightarrow$C1 and
C1$\rightarrow$C0 reversal with direct observation of the resulting
cryptographic or path state. These properties make cryptographic transition,
verification, and reversal explicit parts of the evaluated migration
mechanism.

Other 5GC PQC work includes Koranga~[18] and QORE~[19]. Complementary assurance work includes PQC Validator~[20],
while SCC5G considers a broader PQC-based zero-trust architecture
for critical communication over cellular networks~[21]. In the
present prototype, nghttpx serves as the implementation mechanism
for the deployment-controlled cryptographic boundary. Table~I
summarizes the experimental positioning of the present work relative
to closely related 5GC PQC approaches.

\begin{figure*}[!t]
\centering
\begin{tikzpicture}[
x=1cm,
y=1cm,
>=Latex,
font=\small,
nfbox/.style={draw=BorderBlue,line width=0.9pt,rounded corners=2mm,minimum width=2.55cm,minimum height=1.35cm,align=center},
amfbox/.style={nfbox,fill=NodeAMF},
nrfbox/.style={nfbox,fill=NodeNRF},
bceebox/.style={nfbox,fill=NodeBCEE,draw=AccentTeal,line width=1.0pt,minimum width=2.70cm,minimum height=1.55cm},
hybridinfo/.style={draw=AccentTeal,line width=1.0pt,rounded corners=2mm,fill=NodeHybrid,inner sep=6pt,align=center,text width=5.8cm},
seglabel/.style={font=\footnotesize\bfseries,text=TitleBlue,align=center},
protolabel/.style={font=\scriptsize,text=SoftGray,align=center}
]
\node[amfbox] (amf) at (-7.25,0) {\textbf{Network Function A}\\[-1pt]\footnotesize AMF};
\node[bceebox] (bceea) at (-2.60,0) {\textbf{\BCEEA{}}\\[-1pt]\footnotesize AMF-facing boundary\\[-1pt]\footnotesize \AMFTML{}};
\node[bceebox] (bceeb) at (2.60,0) {\textbf{\BCEEB{}}\\[-1pt]\footnotesize NRF-facing boundary\\[-1pt]\footnotesize \NRFTML{}};
\node[nrfbox] (nrf) at (7.25,0) {\textbf{Network Function B}\\[-1pt]\footnotesize NRF};
\draw[<->,thick,draw=AccentBlue] (amf.east) -- (bceea.west);
\draw[<->,line width=1.5pt,draw=AccentTeal] (bceea.east) -- (bceeb.west);
\draw[<->,thick,draw=AccentBlue] (bceeb.east) -- (nrf.west);
\node[seglabel] at (-5.00,0.48) {$S_1$};
\node[seglabel] at (0,0.48) {$S_2$};
\node[protolabel,text width=2.65cm] at (0,-0.52) {TLS~1.3 / HTTP/2\\[-1pt]\texttt{proto=h2}};
\node[seglabel] at (5.00,0.48) {$S_3$};
\node[hybridinfo] (hybrid) at (0,2.05) {\textbf{Hybrid X25519+ML-KEM-768}\\[3pt]\texttt{X25519MLKEM768}\\[-1pt]\footnotesize NamedGroup 4588};
\draw[<->,densely dotted,line width=0.9pt,draw=AccentTeal] (hybrid.south) -- (0,0.88);
\node[draw=TitleBlue,dash pattern=on 7pt off 4pt,line width=1.15pt,rounded corners=3mm,inner xsep=0.70cm,inner ysep=0.60cm,fit=(bceea)(bceeb)(hybrid)] (cryptoboundary) {};
\node[fill=white,text=TitleBlue,font=\small\bfseries,inner xsep=7pt,inner ysep=2pt] at (cryptoboundary.north) {Deployment-controlled hybrid cryptographic boundary};
\node[protolabel,fill=white,inner sep=1.2pt] at (-5.00,-0.48) {HTTP/2};
\node[protolabel,fill=white,inner sep=1.2pt] at (5.00,-0.48) {HTTP/2};
\end{tikzpicture}
\vspace{2pt}

\begin{minipage}{0.86\textwidth}
\centering
\caption{Deployment-controlled BCEE-mediated AMF--NRF path. Hybrid
X25519+ML-KEM-768 (\HybridGroup{}, \mbox{NamedGroup~4588}) is applied
on the inter-BCEE segment $S_2$; $S_1$ and $S_3$ are NF-facing
compatibility legs.}\label{fig:architecture}
\end{minipage}

\vspace{1pt}
\end{figure*}
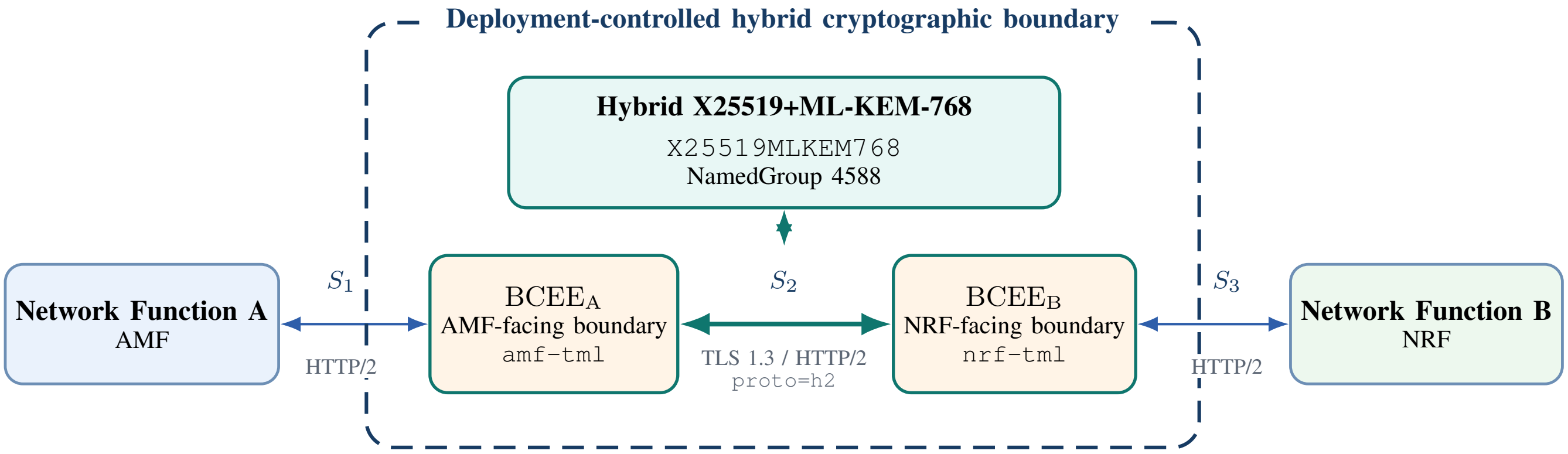

\section{Threat Model and Security Scope}

Fig.~\ref{fig:architecture} depicts the mediated path and its three
segments.
The experiment is motivated by a passive harvest-now-decrypt-later
(HNDL) adversary capable of recording traffic traversing the transport
domain containing the controlled inter-BCEE segment $S_2$ for possible
future cryptanalysis. The evaluated cryptographic property is hybrid
post-quantum/traditional (PQ/T) key establishment on $S_2$. Endpoint
authentication on $S_2$ continues to use the existing classical
certificate-signature mechanism and is outside the present evaluation
scope. Under the passive HNDL model, later compromise of the classical
authentication key does not by itself reveal session secrets derived from
a previously recorded hybrid handshake~[5], [11]. Post-quantum
authentication is a separate migration step.

The mediated path comprises $S_1$: AMF$\leftrightarrow$\BCEEA{},
$S_2$: \BCEEA{}$\leftrightarrow$\BCEEB{}, and $S_3$:
\BCEEB{}$\leftrightarrow$NRF. The BCEEs and the deployment mechanisms
that determine routing and cryptographic policy operate within the trusted
deployment domain. In the laboratory, $S_1$ and $S_3$ are plaintext HTTP
compatibility legs, while the evaluated hybrid cryptographic protection is
introduced on $S_2$. If an NF-facing leg crosses an untrusted transport domain, that leg
requires independent protection---such as native NF TLS where supported,
supplemented by cloud-native workload, network, identity, access-control,
and route-bypass controls---consistent with the 5G security architecture
and procedures in TS~33.501~[22]. Evaluation of these $S_1$ and $S_3$
protection mechanisms is outside the scope of this paper.

Plaintext on $S_1$ and $S_3$ is an experimental brownfield condition, not an
architectural requirement. The participating NF transport profiles are left
unchanged while the evaluated TLS key-establishment profile is controlled
externally on $S_2$. In this paper, ``cryptographic-boundary relocation''
denotes placing enforcement and lifecycle control of that profile at
externally controlled BCEEs, outside the participating NF software stacks.
It does not imply removal or replacement
of cryptographic capabilities natively supported by the NFs. The experiment tests whether this boundary can transition the AMF--NRF
relationship to and from the chosen hybrid profile without modifying
participating NF source code.

For cloud-native deployment, each BCEE should be placed topologically close
to its associated NF so that the NF-facing compatibility leg remains inside a controlled local trust domain. Deployment hardening should combine
workload and network isolation, least-privilege execution, protected
cryptographic material, restricted service exposure, explicit route and
bypass controls, and monitoring of the negotiated cryptographic state. 

For the synchronized C2 capture, the effective $S_2$ group policy is restricted
to X25519MLKEM768, preventing silent negotiation of a classical
key-establishment group in that experiment. A production policy may
additionally permit approved classical groups for compatibility or rollback;
in such deployments, the negotiated group must be observed directly rather
than inferred from successful SBI operation alone. The restriction is an experiment-specific control and does not imply that
all deployment paths must use the hybrid group. Configuration intent is
checked against the observed ClientHello and ServerHello, with the active
policy and configuration provenance recorded for audit. Successful SBI
traffic alone does not establish the negotiated group or the protection
status of the NF-facing legs. Any fallback must therefore be explicit,
authorized, logged, and correlated with the resulting cryptographic state.

\section{System Design}

\subsection{Deployment-Controlled Boundary-Mediated Path}

Without mediation, the selected SBI relationship uses the native
AMF$\leftrightarrow$NRF path. The prototype inserts two
deployment-controlled BCEEs, yielding
AMF$\leftrightarrow$\BCEEA{}$\leftrightarrow$\BCEEB{}$\leftrightarrow$NRF.
OQS-capable nghttpx instances realize the BCEE role in the prototype.
The runtime identifiers \texttt{amf-tml} and \texttt{nrf-tml} denote the
AMF-side and NRF-side Transport Mediation Layer (TML) instances,
corresponding to \BCEEA{} and \BCEEB{}, respectively. This forms the
three segments $S_1$, $S_2$, and $S_3$ defined in Section~III and illustrated in
Fig.~1.

The design objective is deployment-controlled cryptographic policy on
$S_2$. Routing steers the selected SBI relationship through the BCEEs, while
the TLS key-establishment profile on the inter-BCEE segment is configured
independently of the participating NF software stacks. No participating NF
source-code modification is required in the evaluated deployment; routing,
traffic steering, and BCEE cryptographic configuration are changed to
realize the mediated path.

\subsection{Segment Responsibilities and Cryptographic States}

Because both endpoints of $S_2$ are deployment controlled, its TLS
key-establishment profile can be changed without modifying participating
NF source code. The NF-level SBI transaction is preserved across the
mediated path, while the evaluated cryptographic transition is confined to
the inter-BCEE segment.

Three deployment states are evaluated. C0 is the native AMF--NRF path
without BCEE mediation. C1 uses the BCEE-mediated topology with
classical TLS~1.3 X25519 key establishment on $S_2$. C2 retains the same
mediated topology but uses X25519MLKEM768 (X25519+ML-KEM-768) for
hybrid PQ/T key establishment on $S_2$. Thus, the C1/C2 comparison changes the state-specific permitted group
on $S_2$ while holding the mediation topology and OQS-capable
software/provider stack fixed.

The prototype also evaluates manual reversal from C2 to C1 and,
separately, from C1 to C0, with successful SBI operation verified after
each transition. These results demonstrate deployment-controlled cryptographic
transition for the selected AMF--NRF relationship, but they do
not imply transparent operation across all SBA interfaces or
production-scale conditions. BCEE mediation therefore serves as a
brownfield migration control point rather than a replacement for future
native PQC support in 5G NFs.


\section{Experimental Methodology}

\subsection{Prototype Environment}

The virtualized Linux testbed runs an Ubuntu~22.04 VM with 4~vCPUs
and 8~GB RAM on a Windows laptop host with 32~GB RAM and a 1~TB SSD.
It hosts a containerized Open5GS deployment and two OQS-capable nghttpx
instances implementing \BCEEA{} and \BCEEB{}, using the Open Quantum
Safe provider for the evaluated hybrid TLS capability \cite{oqs}. The
prototype retains the fixed OpenSSL~3.0.13 baseline and adds hybrid
capability through a loadable provider rather than replacing the base
platform TLS stack. This mirrors the brownfield condition the paper
addresses, where the cryptographic baseline is constrained by the
deployed environment and its maintenance cycle. It also permits explicit
restriction of the offered group set, which the TLS state-validity
criterion in Section~\ref{subsec:acquisition} requires. OpenSSL~3.5 and later releases natively support \HybridGroup{}, removing
the external oqsprovider dependency for this profile~\cite{oqs}.
Re-baselining onto such a release is next-stage work.

The selected experiment exercises the AMF--NRF SBI relationship. The
NF-facing $S_1$ and $S_3$ legs use plaintext HTTP, whereas $S_2$ uses
TLS~1.3/HTTP/2. The synchronized C2 capture follows
AMF$\rightarrow$\texttt{amf-tml}:7777,
\texttt{amf-tml}$\rightarrow$\texttt{nrf-tml}:8443, and
\texttt{nrf-tml}$\rightarrow$NRF:7777. Configuration specifies the intended group while the captured TLS ClientHello
and ServerHello verify the group actually negotiated. Software and
deployment parameters are summarized in Table~\ref{tab:environment}.

\begin{table}[!t]
\centering
\caption{Prototype Environment}
\label{tab:environment}
\footnotesize
\renewcommand{\arraystretch}{1.10}
\setlength{\tabcolsep}{4pt}
\begin{tabularx}{\columnwidth}{
  >{\raggedright\arraybackslash}p{0.32\columnwidth}
  >{\raggedright\arraybackslash}X
}
\toprule
\textbf{Parameter} & \textbf{Recorded value} \\
\midrule
5G Core          & Containerized Open5GS v2.7.6-131-g782a97e \\
BCEE software    & nghttpx / nghttp2 1.63.0 \\
TLS / PQC stack  & OpenSSL~3.0.13 / oqsprovider 0.12.0-dev \\
$S_2$ profiles   & TLS~1.3/HTTP/2, with X25519 in C1 and \HybridGroup{} in C2 \\
Testbed          & Ubuntu~22.04 VM, 4~vCPUs, 8~GB RAM, Windows laptop host with 32~GB RAM and 1~TB SSD \\
\bottomrule
\end{tabularx}
\end{table}

\subsection{Controlled States and Measurement}

C0 is the native AMF--NRF path and is used only for functional and
reversal validation. C1 and C2 use the same BCEE-mediated topology, with
X25519 and \HybridGroup{}, respectively, on $S_2$. Final performance
inference therefore compares C1 and C2 while holding the mediation
topology fixed. Table~\ref{tab:states} summarizes the two mediated states
used for performance evaluation, including the negotiated group and
key-share sizes that define TLS-state validity.

\begin{table}[!t]
\centering
\caption{Mediated States Used for Performance Evaluation}
\label{tab:states}
\footnotesize
\renewcommand{\arraystretch}{1.10}
\setlength{\tabcolsep}{4pt}
\begin{tabularx}{\columnwidth}{
  >{\centering\arraybackslash}p{0.12\columnwidth}
  >{\raggedright\arraybackslash}X
  >{\centering\arraybackslash}p{0.12\columnwidth}
}
\toprule
\textbf{State} & \textbf{$S_2$ key establishment} & \textbf{N} \\
\midrule
C1 & X25519, NamedGroup~29, 32/32-byte shares & 60 \\
C2 & \HybridGroup{}, NamedGroup~4588, 1216/1120-byte shares & 60 \\
\bottomrule
\end{tabularx}
\end{table}

Complete NF-registration transaction latency is measured at the
AMF-facing boundary.

\begin{equation}
L_{\mathrm{reg}} =
t_{\mathrm{AMF\text{-}face}}(\texttt{201}) -
t_{\mathrm{AMF\text{-}face}}(\texttt{PUT}).
\label{eq:reglatency}
\end{equation}

Both timestamps in Eq.~\eqref{eq:reglatency} are captured from a single
packet-capture instance at the same AMF-facing observation point on the
single-host testbed summarized in Table~\ref{tab:environment}. Each
counted C1/C2 trial creates a fresh $S_2$ TLS connection. Consequently,
the experiment evaluates fresh-connection registration rather than
steady-state connection reuse.

\subsection{Acquisition and Statistical Analysis}
\label{subsec:acquisition}

The final acquisition uses the pre-specified balanced sequence
\texttt{C2-C1-C2-C1-C1-C2-C1-C2-C2-C1-C1-C2}, comprising six separately
initialized blocks per state. The sequence was fixed before acquisition
and was not randomly assigned. Before each block, the requested state is
activated, followed by a 60-s stabilization interval, one uncounted
validation transaction, and ten counted fresh registrations. This yields
120 observations, 60 per state.

For every performance block, a frozen state-specific $S_2$ BCEE
configuration is activated: C1 selects X25519, whereas C2 selects
\HybridGroup{}; the BCEE topology and OQS-capable software/provider
stack are held fixed. The reported deltas therefore reflect system-level
effects of the complete $S_2$ state change and are not isolated KEM costs.

TLS-state validity requires ClientHello/ServerHello \texttt{key\_share}
values matching C1 (NamedGroup~29, 32/32~B) or C2 (NamedGroup~4588,
1216/1120~B). HTTP validity requires the registration \texttt{PUT} and
corresponding \texttt{201 Created}. All 120 protocol-valid observations
are retained. No latency-based exclusion threshold is applied.

Segment-local $S_2$ TLS-establishment timing is obtained from
\texttt{amf-tml} runtime timestamps between \texttt{Connection established}
and \texttt{SSL/TLS handshake completed}. This is a system-level interval,
not a KEM microbenchmark. The $S_2$ TLS-establishment interval and the
complete registration measurement use different observation boundaries and
are therefore reported separately.

The primary performance measure is the difference between the pooled C2
and C1 medians. Because the ten observations within each block share the
same initialized experimental state, confidence intervals are obtained by
resampling complete blocks rather than individual requests. A
30,000-resample whole-block bootstrap with seed~20260828 is used to
report 95\% and 99\% confidence intervals.

Median, mean, SD, IQR, p95, range, and ECDF summarize the latency
distributions. Cliff's delta is reported as a distributional effect size.
An exact two-sided Mann--Whitney $U$ test on the six block medians per
state provides a secondary block-level diagnostic. No request-level
significance test or temporal-order diagnostic is used as a basis for the
primary inference.


\section{Experimental Results}
\label{sec:evaluation}

The results map directly to the research questions. Synchronized
cross-layer validation and additional SBI procedure coverage address RQ1;
packet-level verification of the negotiated hybrid TLS state addresses
RQ2; the C1/C2 timing analyses address RQ3; and the reversal experiments
address RQ4.


\subsection{Synchronized Cross-Layer Validation}

The synchronized C2 capture records a real AMF NF-registration
transaction together with the corresponding hybrid $S_2$ TLS session.
The AMF-originated request carries NF-instance identifier
\SingleRunID{}. After mediation, the NRF-facing \texttt{PUT} preserves
the same NF-instance identifier. The NRF returns \texttt{HTTP 201 Created}
with a \texttt{Location} containing the same identifier, and the successful
response is subsequently observed on the AMF-facing leg.

Matching NF-instance identifiers on the AMF-facing and NRF-facing legs
establish the application transaction across mediation. Transaction
ordering and the known mediated path associate that transaction with the
corresponding $S_2$ TLS connection. Fig.~\ref{fig:amfproof} presents
selected application-layer fields correlating the AMF-originated
registration, the NRF-facing mediated request, and the successful NRF
response.

\begin{figure*}[!t]
\centering

\makebox[0.96\textwidth][c]{%
\begin{minipage}[t]{0.31\textwidth}
\vspace{0pt}
\fbox{%
\begin{minipage}[t][3.35cm][t]{0.91\linewidth}
\scriptsize\raggedright

\textbf{AMF-originated NF registration}\\[-1pt]
\rule{\linewidth}{0.3pt}

{\ttfamily
:method: PUT\\
:path: /nnrf-nfm/v1/\\
\hspace*{1em}nf-instances/\{id\}\\[2pt]
user-agent: AMF
}

\vfill

\textbf{Observed:} AMF-originated NRF
NF-registration request.

\end{minipage}}
\par\vspace{2pt}
{\footnotesize\textbf{(a)} AMF-originated registration}
\end{minipage}%
\hfill
\begin{minipage}[t]{0.31\textwidth}
\vspace{0pt}
\fbox{%
\begin{minipage}[t][3.35cm][t]{0.91\linewidth}
\scriptsize\raggedright

\textbf{NRF-facing mediated request}\\[-1pt]
\rule{\linewidth}{0.3pt}

{\ttfamily
:method: PUT\\
:path: /nnrf-nfm/v1/\\
\hspace*{1em}nf-instances/\{id\}\\[2pt]
user-agent: AMF\\[2pt]
via: 2 nghttpx, 2 nghttpx
}

\vfill

\textbf{Observed:} same NF-instance identifier
on the mediated request.

\end{minipage}}
\par\vspace{2pt}
{\footnotesize\textbf{(b)} NRF-facing mediated request}
\end{minipage}%
\hfill
\begin{minipage}[t]{0.31\textwidth}
\vspace{0pt}
\fbox{%
\begin{minipage}[t][3.35cm][t]{0.91\linewidth}
\scriptsize\raggedright

\textbf{NRF registration response}\\[-1pt]
\rule{\linewidth}{0.3pt}

{\ttfamily
:status: 201 Created\\
location: /nnrf-nfm/v1/\\
\hspace*{1em}nf-instances/\{id\}
}

\par\vspace{6pt}

\textbf{Correlation}\\[-1pt]
\rule{\linewidth}{0.3pt}

AMF \texttt{PUT} with \texttt{\{id\}}\\
$\rightarrow$ mediated \texttt{PUT} with \texttt{\{id\}}\\
$\rightarrow$ NRF \texttt{201 Created}\\
$\rightarrow$ \texttt{Location} with \texttt{\{id\}}

\end{minipage}}
\par\vspace{2pt}
{\footnotesize\textbf{(c)} NRF registration success}
\end{minipage}%
}

\vspace{5pt}

\begin{minipage}{0.84\textwidth}
\centering
\caption{Selected application-layer fields from the synchronized C2
NF-registration capture. The NF-instance identifier, abbreviated as
\texttt{\{id\}}, correlates the AMF-originated request, NRF-facing
mediated request, and successful NRF response.}
\label{fig:amfproof}
\end{minipage}

\end{figure*}


\subsection{Direct Hybrid TLS Negotiation Evidence}

On the synchronized $S_2$ flow, the ClientHello offers
\texttt{0x11EC} (decimal~4588) and carries a 1216-byte
\texttt{key\_exchange}. The ServerHello selects NamedGroup~4588 and carries
a 1120-byte \texttt{key\_exchange}. RFC~10024 assigns
NamedGroup~4588 to \HybridGroup{} \cite{rfc10024}. The observed lengths
match the specified composition: a 1184-byte ML-KEM-768 encapsulation
key and 32-byte X25519 share at the client, and a 1088-byte ML-KEM-768
ciphertext and 32-byte X25519 share at the server.

\begin{figure*}[!t]
\centering

\begin{minipage}[t]{0.495\textwidth}
\vspace{0pt}
\centering
\includegraphics[width=\linewidth]{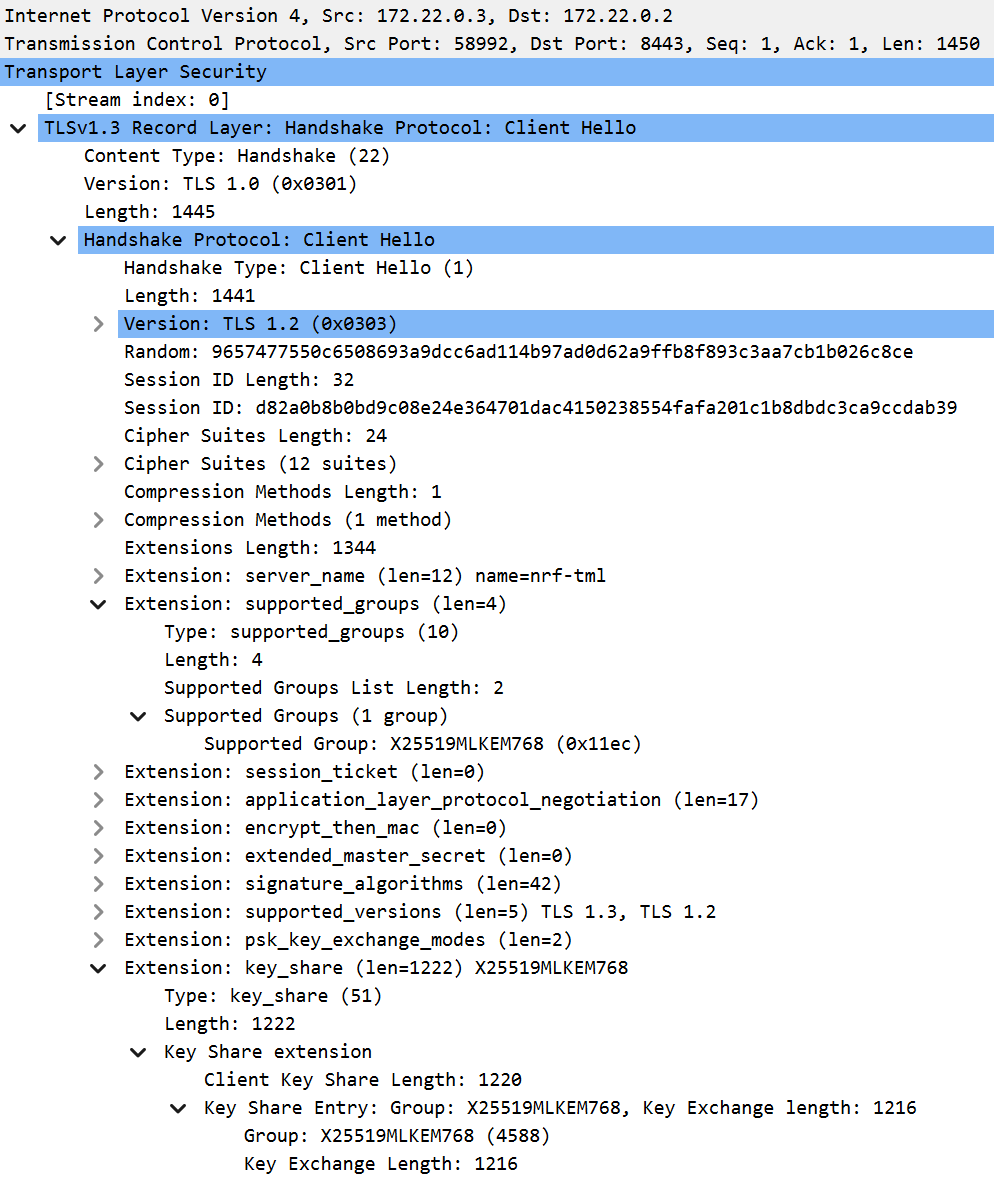}
\vspace{-1pt}
\parbox{\linewidth}{\centering\footnotesize
\textbf{(a) Wireshark ClientHello:} 4588 offered; 1216-byte client share}
\end{minipage}%
\hfill
\begin{minipage}[t]{0.495\textwidth}
\vspace{0pt}
\centering
\includegraphics[width=\linewidth]{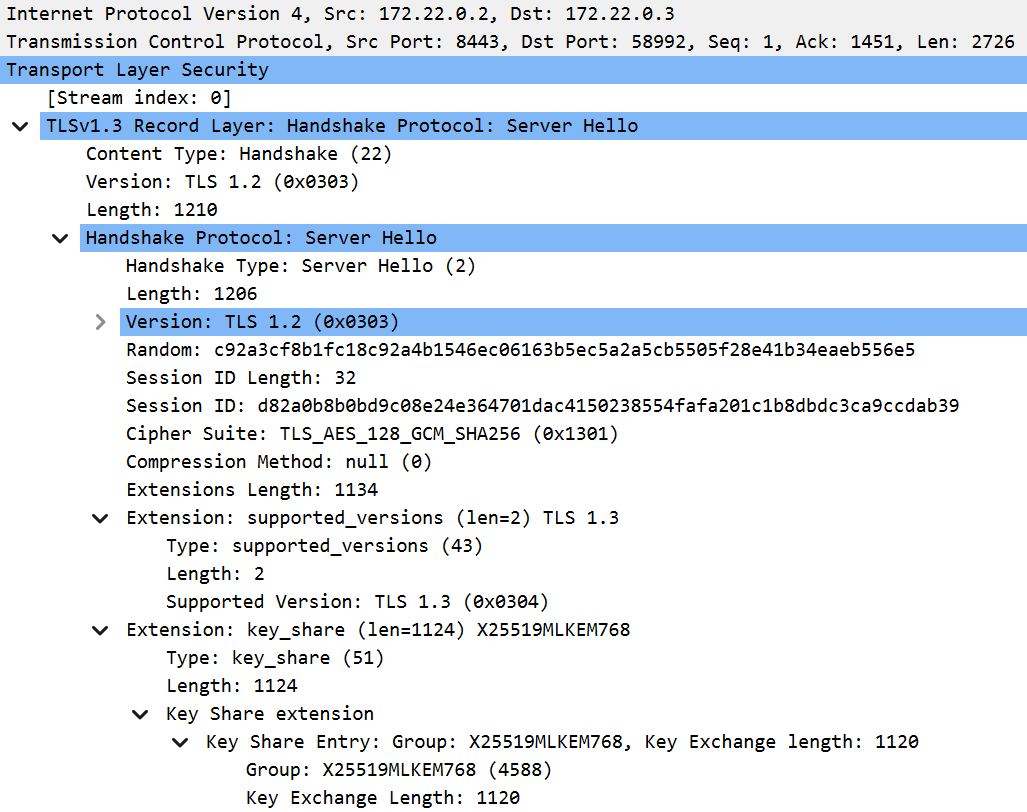}
\vspace{-1pt}
\parbox{\linewidth}{\centering\footnotesize
\textbf{(b) Wireshark ServerHello:} 4588 selected; 1120-byte server share}
\end{minipage}

\vspace{5pt}

\begin{minipage}{0.84\textwidth}
\centering
\caption{Wireshark evidence of hybrid TLS~1.3 negotiation on $S_2$.
The ClientHello offers \HybridGroup{} (NamedGroup~4588) with a
1216-byte key share, and the ServerHello selects the same group with a
1120-byte key share.}
\label{fig:hybridtls}
\end{minipage}

\vspace{-4pt}
\end{figure*}

The OQS-enabled OpenSSL configuration establishes hybrid-group capability,
while the captured ClientHello and ServerHello verify actual negotiation
on the wire. The \texttt{0x0303} value displayed in the protocol tree is the
TLS~1.3 \texttt{legacy\_version} field. TLS~1.3 negotiation is confirmed
by the ClientHello offering and the ServerHello selecting
\texttt{0x0304} through \texttt{supported\_versions}.
Fig.~\ref{fig:hybridtls} presents the corresponding ClientHello and
ServerHello evidence.

The performance captures provide the corresponding validation across all trials.
All 60 C1 trials show NamedGroup~29 with 32/32-byte shares, while all
60 C2 trials show NamedGroup~4588 with 1216/1120-byte shares.

\subsection{Additional SBI Functional Validation}

A separate C2 run extends functional validation beyond the NF-registration
transaction used for synchronized attribution. NRF subscription creation,
NF-resource retrieval, and recurring NF-profile updates also complete
successfully through the mediated path. Repeated \texttt{PATCH} requests
at approximately 10-s intervals receive \texttt{204 No Content},
demonstrating continued NF-management activity after the initial
registration exchange. Representative procedures and responses are
summarized in Table~\ref{tab:functional}.

\begin{table}[!t]
\centering
\caption{Representative SBI Procedures on the Hybrid-Mediated Path}
\label{tab:functional}
\footnotesize
\renewcommand{\arraystretch}{1.10}
\setlength{\tabcolsep}{3.5pt}

\begin{tabularx}{\columnwidth}{
>{\centering\arraybackslash}p{0.15\columnwidth}
>{\raggedright\arraybackslash}X
>{\centering\arraybackslash}p{0.28\columnwidth}
}
\toprule
\textbf{Method} & \textbf{Procedure} & \textbf{Response} \\
\midrule
\texttt{PUT}   & NF registration       & \texttt{201 Created} \\
\texttt{POST}  & NRF subscription      & \texttt{201 Created} \\
\texttt{GET}   & NF-resource retrieval & \texttt{200 OK} \\
\texttt{PATCH} & NF-profile update     & \texttt{204 No Content} \\
\bottomrule
\end{tabularx}

\end{table}


\subsection{Fresh-Connection NF-Registration Latency}

Table~\ref{tab:performance} summarizes all 120 recorded
fresh-connection observations. Complete NF-registration latency has
means of \COneMean{} and \CTwoMean{}~ms and medians of
\COneMedian{} and \CTwoMedian{}~ms for C1 and C2, respectively.
The primary median-based comparison gives a C2--C1 difference of
$+\MedianDelta{}$~ms (\MedianRelativeDelta{}).

These measurements span the complete AMF-facing
\texttt{PUT}$\rightarrow$\texttt{201 Created} transaction, including
mediated path traversal, NRF processing, and return traversal. They
therefore represent system-level transaction completion latency rather
than isolated cryptographic computation.

\begin{figure}[!t]
\centering

\begin{tikzpicture}[x=0.105cm,y=4.35cm,font=\scriptsize,>=Latex]

\draw[draw=SoftGray,->] (5,0) -- (65,0);
\draw[draw=SoftGray,->] (5,0) -- (5,1.07);

\foreach \x in {10,20,30,40,50,60}
  \draw[draw=SoftGray] (\x,0) -- (\x,-0.018)
  node[below,font=\scriptsize]{\x};

\node[anchor=north,font=\scriptsize]
at (35,-0.105) {NF-registration transaction latency (ms)};

\foreach \y/\l in {0/0,0.25/0.25,0.5/0.50,0.75/0.75,1/1}
  \draw[draw=SoftGray] (5,\y) -- (4.2,\y)
  node[left,font=\scriptsize]{\l};
  
\node[rotate=90,font=\scriptsize]
at (-2.2,0.53) {Cumulative probability};

\draw[densely dotted,draw=black!70,line width=0.50pt]
  (5,0.5) -- (64,0.5);
  
\begin{scope}
\clip (5,0) rectangle (64,1.01);

\draw[line width=1.0pt,draw=AccentBlue]
plot[const plot] coordinates {
(8.395123,0.00)
(9.395123,0.0167)
(13.669014,0.0333)
(16.018867,0.0500)
(16.501904,0.0667)
(17.225981,0.0833)
(17.272949,0.1000)
(17.375946,0.1167)
(17.580032,0.1333)
(17.948151,0.1500)
(19.340038,0.1667)
(19.387007,0.1833)
(19.508123,0.2000)
(19.634008,0.2167)
(19.801140,0.2333)
(20.171881,0.2500)
(20.174980,0.2667)
(20.276070,0.2833)
(20.538807,0.3000)
(20.551920,0.3167)
(20.765066,0.3333)
(20.846128,0.3500)
(21.228075,0.3667)
(21.371126,0.3833)
(21.545172,0.4000)
(21.592140,0.4167)
(21.599054,0.4333)
(21.924019,0.4500)
(22.012949,0.4667)
(22.483110,0.4833)
(22.878885,0.5000)
(23.057938,0.5167)
(23.458004,0.5333)
(23.819208,0.5500)
(24.118900,0.5667)
(24.499893,0.5833)
(24.827957,0.6000)
(25.799036,0.6167)
(26.156902,0.6333)
(27.247906,0.6500)
(27.853012,0.6667)
(27.889967,0.6833)
(28.414011,0.7000)
(28.503895,0.7167)
(29.232979,0.7333)
(29.407978,0.7500)
(31.933069,0.7667)
(32.668829,0.7833)
(33.329964,0.8000)
(33.521891,0.8167)
(37.172079,0.8333)
(38.043976,0.8500)
(38.183928,0.8667)
(38.300991,0.8833)
(39.703131,0.9000)
(40.292025,0.9167)
(42.703867,0.9333)
(46.976805,0.9500)
(47.302008,0.9667)
(56.586027,0.9833)
(58.502913,1.0000)
};

\draw[line width=1.0pt,draw=AccentTeal,densely dashed]
plot[const plot] coordinates {
(12.586044,0.00)
(13.586044,0.0167)
(18.172979,0.0333)
(18.801928,0.0500)
(18.951893,0.0667)
(19.264936,0.0833)
(19.881964,0.1000)
(20.740986,0.1167)
(21.937132,0.1333)
(22.258043,0.1500)
(22.721052,0.1667)
(22.820950,0.1833)
(23.216009,0.2000)
(23.234129,0.2167)
(23.307800,0.2333)
(23.457050,0.2500)
(23.468018,0.2667)
(23.522854,0.2833)
(23.597002,0.3000)
(23.628950,0.3167)
(23.848057,0.3333)
(24.095058,0.3500)
(24.394035,0.3667)
(24.456024,0.3833)
(24.883032,0.4000)
(24.926901,0.4167)
(25.403023,0.4333)
(25.537014,0.4500)
(25.573015,0.4667)
(25.725842,0.4833)
(26.315928,0.5000)
(26.386023,0.5167)
(26.482105,0.5333)
(28.004169,0.5500)
(28.944016,0.5667)
(29.567003,0.5833)
(29.633999,0.6000)
(30.683994,0.6167)
(30.926943,0.6333)
(31.865120,0.6500)
(32.854080,0.6667)
(33.041954,0.6833)
(33.928871,0.7000)
(34.115791,0.7167)
(34.910917,0.7333)
(35.292864,0.7500)
(35.444021,0.7667)
(37.399054,0.7833)
(37.652016,0.8000)
(38.937092,0.8167)
(39.664030,0.8333)
(44.062853,0.8500)
(44.339895,0.8667)
(44.414043,0.8833)
(44.929028,0.9000)
(46.344995,0.9167)
(47.940969,0.9333)
(48.364878,0.9500)
(52.371979,0.9667)
(57.424068,0.9833)
(61.221123,1.0000)
};

\end{scope}

\draw[line width=1.0pt,draw=AccentBlue]
(40,0.18) -- (46,0.18);
\node[anchor=west,font=\scriptsize]
at (47,0.18) {C1 X25519};

\draw[line width=1.0pt,draw=AccentTeal,densely dashed]
(40,0.10) -- (46,0.10);
\node[anchor=west,font=\scriptsize]
at (47,0.10) {C2 \HybridGroup{}};

\end{tikzpicture}

\vspace{2pt}

\begin{minipage}{0.88\columnwidth}
\centering
\caption{ECDF of recorded fresh-connection NF-registration transaction
latency for C1 and C2 ($N=60$ per state).}
\label{fig:latency_summary}
\end{minipage}

\end{figure}
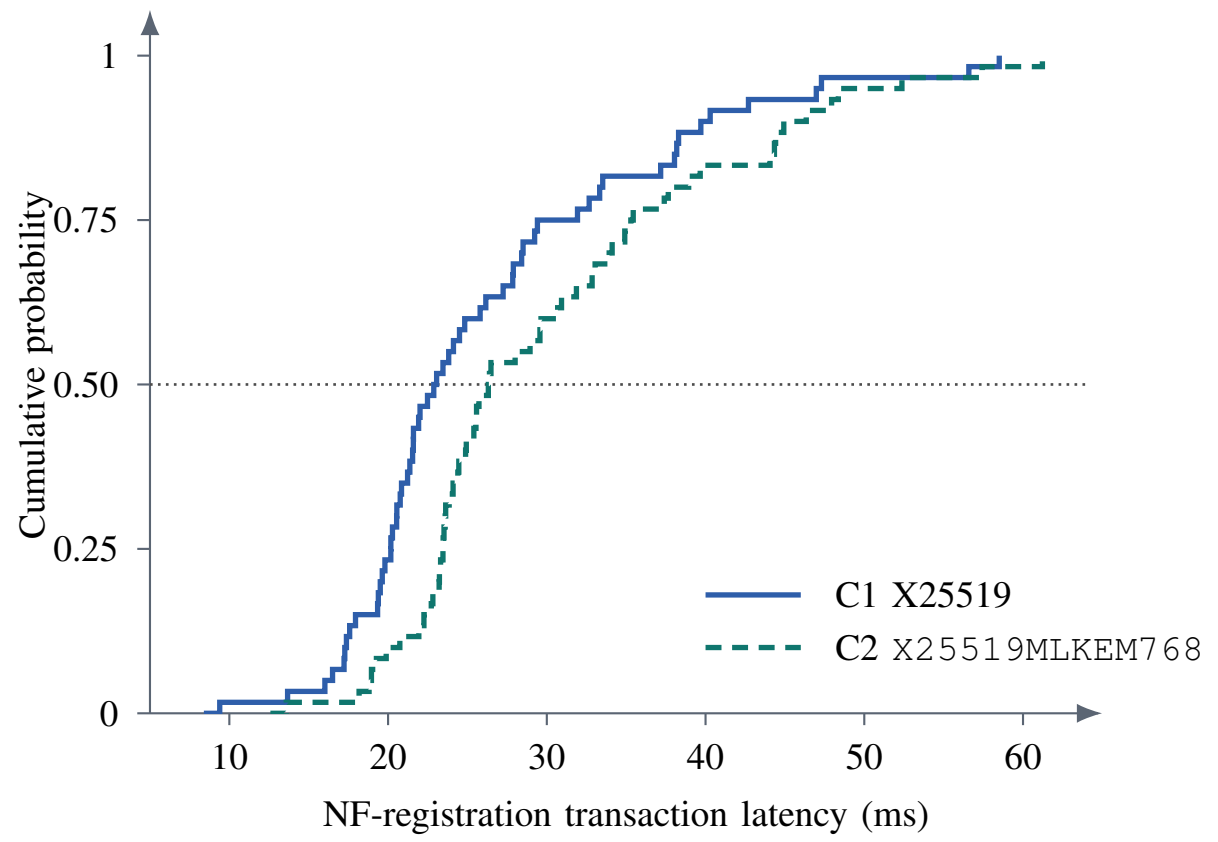

\begin{table}[!t]
\centering
\caption{Fresh-Connection NF-Registration Transaction Latency}
\label{tab:performance}
\footnotesize
\renewcommand{\arraystretch}{1.10}
\setlength{\tabcolsep}{2pt}

\begin{tabularx}{\columnwidth}{
>{\raggedright\arraybackslash}X
>{\centering\arraybackslash}p{0.25\columnwidth}
>{\centering\arraybackslash}p{0.25\columnwidth}
}
\toprule
\textbf{Statistic} &
\textbf{C1} &
\textbf{C2} \\
\midrule
N & 60 & 60 \\
Mean $\pm$ SD & \COneMean{} $\pm$ \COneSD{} &
\CTwoMean{} $\pm$ \CTwoSD{} \\
Median & \COneMedian{} & \CTwoMedian{} \\
Q1--Q3 & \COneQOne{}--\COneQThree{} &
\CTwoQOne{}--\CTwoQThree{} \\
p95 & \COnePNinetyFive{} & \CTwoPNinetyFive{} \\
Min--Max & \COneMin{}--\COneMax{} & \CTwoMin{}--\CTwoMax{} \\
\midrule
\multicolumn{3}{>{\raggedright\arraybackslash}p{0.91\columnwidth}}{%
Median C2--C1 difference: $+\MedianDelta{}$~ms
(\MedianRelativeDelta{}).} \\
\multicolumn{3}{>{\raggedright\arraybackslash}p{0.91\columnwidth}}{%
Whole-block bootstrap 95\% CI: \MedianBlockCI{}~ms.} \\
\multicolumn{3}{>{\raggedright\arraybackslash}p{0.91\columnwidth}}{%
Whole-block bootstrap 99\% CI: \MedianBlockCINinetyNine{}~ms.} \\
\bottomrule
\end{tabularx}
\vspace{1pt}
\begin{minipage}{0.96\columnwidth}
\scriptsize
All values are in ms. C1 uses X25519 and C2 uses \HybridGroup{}.
Confidence intervals use whole-block bootstrap resampling.
\end{minipage}
\end{table}


\subsection{Block-Aware Inference and Robustness}

For the complete NF-registration transaction, the pooled C2--C1 median
difference is $+\MedianDelta{}$~ms (\MedianRelativeDelta{}).
Resampling complete state blocks gives a 95\% whole-block bootstrap CI
of \MedianBlockCI{}~ms and a 99\% CI of
\MedianBlockCINinetyNine{}~ms. Both intervals exclude zero, supporting
a positive C2--C1 fresh-registration latency shift at the AMF-facing
measurement boundary.

Cliff's delta is $\CliffsDelta{}$, indicating a distributional shift
toward higher C2 latency. An exact two-sided Mann--Whitney $U$ test on
the six block medians per state gives $U=\BlockMWU{}$, $p=\BlockMWP{}$,
consistent with the block-aware bootstrap result.

The pre-specified sequence is balanced across the 12 acquisition
positions, with six separately  initialized blocks per state and ten
counted fresh registrations per block. The inference therefore rests on
the whole-block bootstrap intervals and the block-level Mann--Whitney
check, while avoiding a request-level independence assumption.
Distributional overlap nevertheless remains visible in
Fig.~\ref{fig:latency_summary}.


\subsection{Timing Localization Across the Fresh-Connection Path}
\label{subsec:timinglocalization}

To examine where C1/C2 timing separation appears, the same 12-block
acquisition is reported at two observation boundaries: $S_2$ TLS
establishment and completion of the NF-registration transaction.

\begin{table}[!t]
\centering
\caption{Timing Across the Fresh-Connection Path}
\label{tab:timinglocalization}
\footnotesize
\renewcommand{\arraystretch}{1.10}
\setlength{\tabcolsep}{1.6pt}
\begin{tabularx}{\columnwidth}{
  >{\raggedright\arraybackslash}X
  >{\centering\arraybackslash}p{0.12\columnwidth}
  >{\centering\arraybackslash}p{0.12\columnwidth}
  >{\centering\arraybackslash}p{0.12\columnwidth}
  >{\centering\arraybackslash}p{0.24\columnwidth}
}
\toprule
\textbf{Boundary} & \textbf{C1} & \textbf{C2} &
\textbf{$\Delta$} & \textbf{95\% CI} \\
\midrule
$S_2$ TLS establishment & \STwoCOneMedian{} & \STwoCTwoMedian{} &
$+$\STwoDelta{} & \STwoCI{} \\
Complete registration ($S_1$) &
\COneMedian{} & \CTwoMedian{} & $+$\MedianDelta{} & \MedianBlockCI{} \\
\bottomrule
\end{tabularx}
\vspace{1pt}
\begin{minipage}{0.96\columnwidth}
\scriptsize
All values are in ms. $\Delta=\mathrm{median}(C2)-\mathrm{median}(C1)$.
Confidence intervals use whole-block bootstrap resampling. The two intervals
use different observation boundaries and are therefore reported separately.
\end{minipage}
\end{table}

Table~\ref{tab:timinglocalization} compares the two timing boundaries.
Median $S_2$ TLS-establishment time increases from
\STwoCOneMedian{}~ms in C1 to \STwoCTwoMedian{}~ms in C2. The
complete NF-registration median increases from \COneMedian{} to
\CTwoMedian{}~ms. These changes correspond to C2--C1 differences of
$+\STwoDelta{}$~ms (\STwoRelativeDelta{}) and
$+\MedianDelta{}$~ms, respectively. Both 95\%
whole-block confidence intervals exclude zero. For the $S_2$
difference, the 99\% interval is \STwoCINinetyNine{}~ms.

The separation is especially clear at $S_2$. The six C1 block medians
span \STwoCOneBlockRange{}~ms, compared with
\STwoCTwoBlockRange{}~ms for C2. All six C2 values are above the C1
values. An exact two-sided Mann--Whitney $U$ test gives
$U=\STwoBlockMWU{}$ and $p=\STwoBlockMWP{}$. This is the maximum
possible $U$ statistic for six blocks per state.

A positive C2--C1 difference is therefore observed at both reported
timing boundaries. The complete registration measurement captures
completion of the fresh-registration transaction at the AMF-facing
boundary, whereas the $S_2$ measurement captures segment-local TLS
establishment at the controlled inter-BCEE boundary. These intervals are
read separately rather than as additive components, because neither
isolates a single implementation cost.

Together, Fig.~\ref{fig:latency_summary} and
Table~\ref{tab:timinglocalization} give the final timing check for the
fresh-connection experiment. The ECDF shows the complete-registration
latency distribution, while the timing table reports the controlled
$S_2$ TLS-establishment boundary from the same 12-block run. Read
together, they show a consistent C2--C1 increase without treating the two
timings as the same measurement.

This interpretation keeps the result conservative. The registration value is
the measured outcome of the complete mediated transaction, while the $S_2$
value identifies the controlled TLS-establishment boundary. The analysis does
not decompose proxy processing, NRF processing, queueing, or return traversal
into separate component costs.

\subsection{Manual Reversal and Native-Path Restoration}

The C2$\rightarrow$C1 run first shows NamedGroup~4588 and then
X25519 NamedGroup~29. Once C1 is active, NF registration completes with
a \texttt{201 Created} response. A separate C1$\rightarrow$C0 run
records NamedGroup~29 before native-path restoration. After the native path is restored,
no frames use port~8443. The AMF instead reaches the NRF directly on
port~7777, where registration returns \texttt{201 Created} and later
\texttt{PATCH} requests receive \texttt{204 No Content}. Each reversal
was exercised once, so these observations support operational
reversibility rather than provide transition-time statistics.

\section{Discussion}
\label{sec:discussion}

The C1/C2 comparison changes the cryptographic profile on $S_2$ while
keeping the mediated path fixed. This makes the timing result easier to
interpret. Positive C2--C1 differences appear at both fresh-connection
measurement boundaries, $S_2$ TLS establishment and complete
NF-registration completion. In both cases, the whole-block 95\%
confidence interval excludes zero.

These two measurements answer different parts of the question. The
$S_2$ result shows timing separation at the controlled inter-BCEE TLS
boundary. The complete-registration result shows the effect visible at
the AMF-facing transaction boundary. They should be read together, but
not added together. Neither value is an isolated ML-KEM implementation cost. Both are system-level runtime measurements of the evaluated system.

The main migration value is that BCEE mediation gives the deployment a
controlled place to change the TLS key-establishment profile. The
participating NFs keep their source code unchanged, while $S_1$ and
$S_3$ preserve compatibility in this laboratory deployment. This is the
brownfield advantage. Cryptographic change can move at the boundary
without waiting for every NF software stack to move at the same time.

The packet evidence matters because the service success alone does not prove
which cryptographic state was negotiated. The synchronized C2 capture
links a real NF-registration transaction to the corresponding $S_2$ TLS
session and verifies \HybridGroup{} as NamedGroup~4588. The same
NF-instance identifier is preserved across the mediated request and the
successful response. Additional \texttt{PUT}, \texttt{POST},
\texttt{GET}, and recurring \texttt{PATCH} procedures show that SBI
activity continues through the mediated path.

The evaluation scope is deliberately focused. It studies one AMF--NRF
relationship on a single virtualized host where both NFs and both BCEEs
share four vCPUs. The reported latencies are best read as comparative
system measurements for this deployment, rather than as production
benchmarks.

The same method can now be extended in a direct way. Next-stage work
should test additional NF pairs and SBI procedures, connection reuse,
CPU and memory cost, different BCEE placements, failover behavior,
post-quantum authentication, and automated crypto-state monitoring.
Together, these extensions would move the present single-relationship
demonstration toward a broader crypto-agile 5G Core deployment model.

\section{Conclusion}

The experiments show that a brownfield 5G Core path can be
cryptographically segmented so that the inter-BCEE $S_2$ profile can be
changed independently between classical X25519 and hybrid
X25519MLKEM768 operation, without modifying participating NF source code.
This creates a practical migration boundary in which cryptographic change
can be introduced independently of NF software evolution.

Synchronized packet evidence links a real AMF NF-registration transaction
to the corresponding $S_2$ TLS session and directly verifies negotiation
of hybrid NamedGroup~4588. Successful \texttt{PUT}, \texttt{POST}, \texttt{GET}, and recurring
\texttt{PATCH} procedures further show that the evaluated SBI procedures
remain operational under the hybrid configuration. The evidence connects application-level operation with the
cryptographic state actually negotiated on the wire.

The balanced 12-block evaluation comprising 120 fresh NF registrations
shows positive C2--C1 timing differences at both the AMF-facing
registration boundary and the $S_2$ TLS-establishment boundary. The
reported median differences are +3.383~ms and +4.599~ms, respectively,
with whole-block 95\% confidence intervals excluding zero at both
boundaries.

The manual C2$\rightarrow$C1 rollback and the separate
C1$\rightarrow$C0 native-path restoration add an operational dimension
to the migration model. They demonstrate that the
negotiated cryptographic profile can be returned from hybrid to classical
operation and that the mediated path can separately be restored to native
AMF--NRF operation. In both cases, the resulting state is directly visible
in packet evidence and successful SBI operation follows the transition.

Taken together, segmented cryptographic control, standardized
wire-verified hybrid negotiation, representative  SBI functionality, block-aware performance
evaluation, controlled reversibility, and crypto-state observability form
a practical experimental framework for brownfield 5G Core crypto-agility.
The same framework provides a direct path toward broader multi-SBI
validation, automated cryptographic-state control, and operationally
observable post-quantum migration across larger 5G Core deployments.

\balance

\end{document}